\documentclass[aps,prd,twocolumn,showpacs,nofootinbib]{revtex4-1}

\usepackage{graphicx,color}

\begin{document}

\title{The $B_c (B_c^*)$ meson production via the proton-nucleus and the nucleus-nucleus collision modes at the colliders RHIC and LHC}

\author{Gu Chen$^1$}\email{email: speecgu@gzhu.edu.cn}
\author{Chao-Hsi Chang$^{2,3}$}\email{email: zhangzx@itp.ac.cn}
\author{Xing-Gang Wu$^4$}\email{email: wuxg@cqu.edu.cn}

\address{$^1$School of Physics $\&$ Electronic Engineering, Guangzhou University, Guangzhou 510006, P.R. China.\\
$^2$Institute of Theoretical Physics, Chinese Academy of Sciences, P.O.Box 2735, Beijing 100080, P.R. China.\\
$^3$ School of Physical Sciences, University of Chinese Academy of Sciences, Beijing 100049, China\\
$^4$Department of Physics, Chongqing University, Chongqing 401331, P.R. China. }

\date{\today}

\begin{abstract}

In the paper, we make a comprehensive study on the hadroproduction of the $B_c (B_c^*)$ meson via the gluon-gluon fusion mechanism at the RHIC and LHC colliders. Total and differential cross sections via the proton-nucleus ($p$-N) and nucleus-nucleus (N-N) collision modes have been discussed under various collision energies. To compare with those via the proton-proton collision mode at the LHC, we observe that sizable number of $B_c (B_c^*)$-meson events can also be produced via the $p$-N and N-N collision modes at the RHIC and LHC. If assuming the spin-triplet $B^*_c$ meson directly decays to the spin-singlet $B_c$ meson with $100\%$ probability, $1.2 \times 10^5$ and $4.7 \times 10^5$ $B_c$-meson events can be produced via the $p$-Au and Au-Au collision modes at the RHIC in one operation year; $5.8 \times 10^6$ and $4.6 \times 10^6$ $B_c$-meson events can be produced via the $p$-Pb and Pb-Pb collision modes at the LHC in one operation year. \\


\end{abstract}

\maketitle

\section{Introduction}

The $(c\bar{b})$-quarkonium is an important system for understanding various aspects of Quantum Chromo-dynamics (QCD). It carries flavors explicitly and decays via the weak interaction only, which has a relatively longer lifetime than any other doubly heavy quarkoniums. The masses of the $c$-quark and the $\bar{b}$-quark as well as the relevant CKM matrix elements happen to cause comparable decay rates for either the constituent $c$-quark decay channels or the constituent $\bar{b}$-quark decay channels. Those properties make the $(c\bar{b})$-quarkonium ($B_c$ and $B_c^*$ etc) a fruitful ``laboratory" for testing the QCD potential models and for understanding the weak decay mechanism of the two heavy flavors simultaneously.

It is known that the behavior of the formation and dissociation of the $(c\bar{c})$-quarkonium ($J/\psi$ etc) in heavy ion collisions offers important information about the quark-gluon plasma (QGP) produced in high-energy heavy ion collisions. From lattice calculations and potential models, the binding energy of the $(c\bar{b})$-quarkonium is greater than that of the $(c\bar{c})$-quarkonium (but lower than that of the $(b\bar{b})$-quarkonium), thus it is believed that the behavior of the formation and dissociation of the $(c\bar{b})$-quarkonium in high-energy heavy ion collisions will offer more useful information about the QGP to complement those via the formation and dissociation of the $(c\bar{c})$-quarkonium, e.g. the dissociation temperature of the $(c\bar{b})$-quarkonium in the QGP must be higher than that of the $(c\bar{c})$-quarkonium, so the studies on QGP via the $(c\bar{b})$-quarkonium in high-energy heavy ion collisions, paralleled with those via the $(c\bar{c})$-quarkonium, are important and interesting~\cite{zhuangpf1}. Aiming this in mind, we think that as a preliminary step, the direct production of the $(c\bar{b})$-quarkonium in high-energy heavy ion collisions should be well understood. In the paper, we shall restrict ourselves to study the direct production mechanism of the $(c\bar{b})$-quarkonium in high-energy heavy ion collisions.

According to the Non-Relativistic Quantum Chromo-Dynamics (NRQCD) effective theory~\cite{Bodwin:1994jh}, the $(c\bar{b})$-pair in the quarkonium could be either in color-singlet or in color-octet state which can form the measurable color-singlet state by properly grabbing soft gluons or light quarks. In the paper, as a leading-order estimation of the $(c\bar{b})$-quarkonium hadroproduction via the nucleus collision modes, we shall concentrate our attention on the color-singlet $B_c(|c\bar{b}[^1S_0]\rangle)$ and $B_c^*(|c\bar{b}[^3S_1]\rangle)$, which are in spin-singlet and spin-triplet states, respectively. The $B_c^*$ meson can decay to the ground state $B_c$ meson with almost $100\%$ possibility via electromagnetic interactions, and the production of $B_c^*$ can be an additional source of the $B_c$ meson production.

For the production of color-singlet bound-state, the production rates of the $B_c$ meson can be factorized into the convolution of the perturbative short-distance coefficient and the non-perturbative wavefunction at the origin of the bounding system~\cite{Chang:1979nn}. Initial theoretical studies on the $B_c$ meson production at the $e^+ e^-$ collider had been given in Refs.\cite{Chang:1991bp, Chang:1992bb, Kiselev:1994qp}, which stimulated the first experimental search of the $B_c$ meson by the collaborations such as OPAL~\cite{Ackerstaff:1998zf}, DELPHI~\cite{Abreu:1996nz} and ALEPH~\cite{Barate:1997kk} at the Large Electron Positron Collider (LEP)~\cite{Assmann:2002th}. But due to small production rate and low integrated luminosity accumulated at the LEP-I and LEP-II runs, only very few candidate events had been observed by those collaborations, so they could not ensure its observation. However in future high luminosity $e^+e^-$ and the $ep$ colliders, such as the International Linear Collider (ILC)~\cite{Djouadi:2007ik}, the super Z-factory~\cite{Erler:2000jg, zf2} and the Large Hadron Electron Collider (LHeC)~\cite{AbelleiraFernandez:2012cc}, sizable number of $B_c$ meson events may be generated via the electro-production or the photo-production mechanisms~\cite{Chang:2010am, Yang:2011ps, Yang:2013vba, Sun:2013wuk, Zheng:2015ixa, Zheng:2017xgj, Bi:2016vbt}.

In the same period of time, suggestions for observing the $B_c$ meson at the TEVATRON and the Large Hadron Collider (LHC) were investigated in the literature~\cite{Chang:1992jb, Chang:1994aw, Kolodziej:1995nv, Berezhnoy:1994ba, Chang:1996jt, Berezhnoy:1996ks, Baranov:1997sg, Baranov:1997wy}. In 1998, the CDF collaboration at the TEVATRON reported their first observation of the $B_c$ meson~\cite{Abe:1998wi} by using the ``golden channels" such as the $B_c\to J/\psi$ semileptonic decays~\cite{Lusignoli:1990ky, Chang:1992pt, Scora:1995ty}. This observation was subsequently confirmed by more precise data accumulated at the TEVATRON and the LHC. Those experimental progresses aroused people's wide interests on the $B_c$ meson. Later on, more hadroproduction mechanisms, in addition to the dominant gluon-gluon fusion mechanism, have also been discussed in the literature~\cite{Chang:2003cr, Chang:2004bh, Chang:2005wd, Chang:2005bf}. Those works were culminated in a generator BCVEGPY~\cite{Chang:2003cq, Chang:2005hq, Chang:2006xka, Wu:2013pya, Chang:2015qea}, which is now widely accepted by various experimental collaborations for simulating the $B_c$ meson events at the hadronic colliders.

The ALICE at the LHC and the Star etc at the RHIC are heavy-ion detectors, which may work in proton-nucleus ($p$-N) mode or in nucleus-nucleus (N-N) collision mode. Roughly, for the same integrated luminosity, one would expect that two or more orders larger cross sections for the $B_c$-meson production can be achieved via heavy-ion collisions than those of the $pp$ collisions. As will be shown later, about $10^6$ $B_c^{(*)}$-meson events can be produced in Star or ALICE with their designed luminosity in one operation year. Thus, even considering the direct production mechanisms only, in addition to the $pp$ collision at the LHC, those two operating heavy-ion colliders can hopefully be important plateau for studying the $B_c$-meson properties as long as the signals can be separated from the background.

At present, the charmonium and bottomonium productions via the $p$-N or N-N collision mode have been analyzed theoretically~\cite{Arleo:2012rs, Fujii:2013gxa, Ma:2015sia, Ducloue:2015gfa, Lansberg:2016deg, Vogt:2010aa} and experimentally~\cite{Adare:2013ezl, Abelev:2013yxa, Aaij:2013zxa, Abelev:2014zpa, Adam:2015iga, Adam:2015jsa, Aad:2015ddl, Aaij:2016eyl, Adam:2016ohd, Adare:2016psx, Sirunyan:2017mzd, Aaij:2017cqq}. A comparative agreement between the predictions and the measurements have been achieved. Studies on the $B_c$ meson production via the $p$-N and N-N collision modes are still at the initial stage, and it is important to study the availability on the $B_c (B_c^*)$-meson production via the $p$-N and N-N collision modes. The $B_c$-meson production via heavy-ion collision provides a brand-new perspective insight into the $B_c$ meson hadroproduction mechanism. It offers a crucial supplement to various aspects of QCD such as to achieve useful insights into the heavy ion structure. The $B_c$ meson hadroproduction at the RHIC and LHC in the case of $p$-N or N-N collision at the RHIC and LHC shall be affected by extra effects from the incident nucleus~\cite{Vogt:2010aa, Andronic:2015wma, Lansberg:2015uxa}, since the participant nucleon (proton or neutron) are bound in the incident nucleus. Phenomenologically, before collision the participant partons may loose energy in the way through the nucleus shadowing effect, and after collision the newly produced $B_c$ meson may be destroyed and/or absorbed when it travels through the rest part of the nucleus. Moreover, the physical effects emerging in the $B_c$ production are also expected to improve our understanding of the nucleus parton distribution functions.

The remaining parts of the paper are organized as follows. In Sec.II, we focus on the direct production mechanism for the $B_c$ meson production in Star etc (RHIC) and in ALICE (LHC) via the dominant gluon-gluon fusion mechanism and present the calculation technology~\footnote{The $B_c$ meson production via the extrinsic or intrinsic heavy quark mechanism is in preparation, which is also important and sizable in small and intermediate $p_t$ regions.}. In Sec.III, numerical results for the production cross-sections together with their uncertainties under suitable choices of input parameters are presented, and those for the $pp$ collision mode are also presented for comparison. Sec.IV is reserved for a summary.

\section{Calculation Technology}

For the leading-order calculation, which is at the $\alpha_s^4$-order level, the gluon-gluon fusion mechanism via the subprocess, $g+g \to B_c+ \bar{c}+b$, provides dominant contribution to the hadroproduction of $B_c$ meson.

\begin{figure}[htb]
\includegraphics[width=0.4\textwidth]{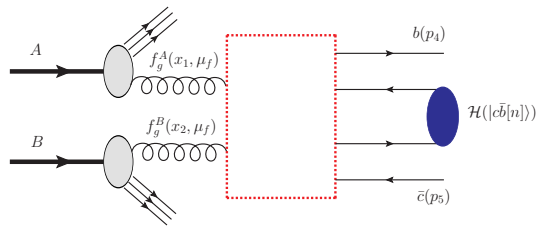}
\caption{The schematic factorization picture for the hadronic production of $B_c$ meson via the gluon-gluon fusion mechanism, where the dashed box stands for the hard interaction kernel, which is perturbatively calculable. }
\label{feyn}
\end{figure}

As the first step, two heavy quark pairs, ($c\bar{c}$) and ($b\bar{b}$), are produced via hard scattering process, $g+g \to c+ \bar{c}+b+\bar{b}$; it is pQCD calculable since the intermediate gluon should be hard enough to form either a ($c\bar{c}$)-pair or a ($b\bar{b}$)-pair. Then the $c$-quark and the $\bar{b}$-quark shall be hadronized into the $B_c$ meson, $c+ \bar{b}\to B_c$, via a non-perturbative way which can be characterized by the wave function at the origin. The schematic factorization picture for the hadronic production of $B_c$ meson via the gluon-gluon fusion mechanism is shown in Fig.(\ref{feyn}).

By using the NRQCD factorization theorem, we can write the total production cross-section as
\begin{widetext}
\begin{eqnarray}\label{corsec}
d\sigma_{AB \to {\cal H}(|c\bar{b}[n]\rangle)+X} &=& N_A N_B \int dx_1 dx_2 f^{\rm A}_{g}(x_1, \mu_{f}) f^{\rm B}_{g}(x_2, \mu_{f}) d\hat\sigma_{gg \to |c\bar{b}[n]\rangle+ X} \langle{\cal O}^{\cal H}[n] \rangle\;,
\end{eqnarray}
\end{widetext}
where $\langle{\cal O}^{\cal H}[n] \rangle$ is the long-distance matrix element, which is proportional to the inclusive transition probability of the perturbative state $|c\bar{b}[n]\rangle$ into the quarkonium state ${\cal H}(|c\bar{b}[n]\rangle)$. The symbols $A$ and $B$ stand for $p$ or N for the incident hadron to be proton or nucleus, respectively. For the nucleus gluon density, we adopt the general approximation that the gluon PDFs in proton and neutron are the same, thus there are overall factors $N_A$ and $N_B$ in the formulae. $N_A$ or $N_B$ is the nuclear number in the incident nucleus, e.g., $N_{\rm Au}=197$ for the gold nucleus ($^{197}_{79}\rm Au$), and $N_{\rm Pb}=208$ for the lead nucleus ($^{208}_{82}\rm Pb$). The PDFs $f_g^{\rm A}(x_1, \mu_{f})$ and $f_g^{\rm B}(x_2, \mu_{f})$ are gluon PDFs inside the nucleon bound in the nucleus $A$ or $B$ accordingly, which carry the fractions $x_1$ or $x_2$ of the nucleon momentum at the factorization scale $\mu_f$. For the $pp$ collision mode, we have $N_A=N_B=1$, and the PDFs are reduced to the usual gluon PDFs inside the free proton as $f_g^{p}(x_{1,2}, \mu_{f})$. Furthermore, we need to consider the collision geometry and the spatial dependence of the shadowing parameterization effect~\cite{Aubert:1987da, Gavin:1990gm}. In our present calculation, we shall adopt the nCTEQ15 version~\cite{Kovarik:2015cma} as the nucleus PDF, which incorporate those effects into the PDF via global fit of the experimental data.

By ignoring the small spin-splitting effect, the color-singlet $S$-wave matrix element $\langle{\cal O}^{\cal H}[^1S_0] \rangle=\langle{\cal O}^{\cal H}[^3S_1] \rangle={|R_S(0)|^2}/{4\pi}$, where the radial wavefunction at the origin $|R_S(0)|^2=1.642\;{\rm GeV}^3$ under the Buchmuller-Tye potential model~\cite{Eichten:1994gt, Eichten:1995ch}. $d\hat\sigma_{gg \rightarrow |c\bar{b}[n]\rangle + X}$ is the differential cross-section of the hard subprocess,
\begin{eqnarray}\label{hardc}
&& d\hat\sigma_{gg \to |c\bar{b}[n]\rangle + X} = \frac{1}{2 x_1 x_2 S_{AB}} \overline{\sum}  |{\cal M}|^{2} d\Phi_m \;.
\end{eqnarray}
Here $S_{AB}$ is the center-of-mass energy of the incident hadrons $A$ and $B$. $\overline{\sum}$ means the averaging over the spin states of the
incident gluons and summing over the color and spin of all final particles. ${\cal M}$ is the hard scattering amplitude for the subprocess $g(k_{1})+g(k_{2}) \to |(c\bar{b})_{\bf 1}[n]\rangle(p_{3})+b(p_{4})+\bar{c}(p_{5})$. $d\Phi_m$ is the $m$-body phase space defined as,
\begin{equation}
d{\Phi _m} = {(2\pi )^4} {\delta ^4}({k_1} + {k_2} - \sum\limits_f^m {{q_f}} )\prod\limits_{f = 1}^n {\frac{d\vec{q}_{f}} {(2\pi)^3 2q_f^0}} .
\end{equation}
We adopt the generator BCVEGPY to deal with the hard scattering amplitude and the phase space.

\section{Numerical results}\label{results}

\subsection{Input parameters}

As for the heavy quark masses, we take $m_c=1.5$ GeV and $m_b=4.9$ GeV. To ensure the gauge invariance of the hard scattering amplitude, the $B_c$ meson mass is the sum of the two constituent quark masses, $M_{B_c}=m_b+m_c$. The renormalization scale and the factorization scale are set to be the same and are taken as the transverse mass of the $B_c$ meson, i.e., $\mu_R=\mu_f=M_t=\sqrt{M_{B_c}^2+p_t^2}$.

In the $p$-N and N-N collision modes, the  nucleus beam energy per nucleon $E_{\rm N}$ is related to the proton beam energy $E_p$ and the charge-to-mass ratio of the incoming nucleus $Z/N_A$, $E_{\rm N}=E_p Z/N_A$~\cite{Vogt:2010aa}. 

At the LHC, the collision energy for its $pp$ collision mode is $\sqrt{S_{pp}}=13$ TeV and the proton beam energy $E_p=6.5$ TeV; the collision energy for its $p$-Pb collision model is $\sqrt{S_{p\rm Pb}}=\sqrt{4E_pE_{\rm Pb}}= 8.16$ TeV; the collision energy for its Pb-Pb collision mode is $\sqrt{S_{\rm PbPb}}=5.02\,\rm TeV$. At the RHIC, the collision energy is taken as $\sqrt{S_{p\rm Au}}=0.2$ TeV for the $p$-Au collision and $\sqrt{S_{\rm AuAu}}=0.2$ TeV for the Au-Au collision, respectively.

\subsection{Basic results}

\begin{table*}[htb]
\begin{tabular}{|c|c|c|c|c|c|c|c|c|}
\hline
& \multicolumn{2}{|c|}{RHIC} & \multicolumn{3}{|c|} {LHC}\\
\hline
& $p$-Au (0.2 TeV)  & Au-Au (0.2 TeV)  & $pp$ (13 TeV)   & $p$-Pb (8.16 TeV)  & Pb-Pb (5.02 TeV) \\
\hline
$\sigma_{B_c}$ (nb)~& $8.19$ & $1.76\times 10^3$ & $4.24\times 10^1$    & $3.29\times 10^3$ & $3.69\times 10^5$ \\
\hline
$\sigma_{B_c^*}$ (nb)~& $1.93\times 10^1$  & $4.15\times 10^3$ & $1.05\times10^2$   & $8.26\times 10^3$ & $9.21\times 10^5$ \\
\hline
$\sigma_{|(^1S_0)_{\bf 8}g\rangle}$ (nb)& $1.82\times 10^{-1}$ & $3.94\times 10^1$ & $7.89\times 10^{-1}$ & $5.96\times 10^1$ & $6.83\times 10^3$ \\
\hline
$\sigma_{|(^3S_1)_{\bf 8}g\rangle}$ (nb)& $8.36\times 10^{-1}$  & $1.83\times 10^2$ & 3.40 & $2.55\times 10^2$ & $2.90\times 10^4$ \\
\hline
\end{tabular}
\caption{Total cross sections (in unit: nb) for the $B_c^{(*)}$-meson production via the gluon-gluon fusion mechanism at the RHIC and the LHC, respectively. Typical collision energies for various collision modes are adopted. As a comparison, we present the cross-sections for the two color-octet $|(^1S_0)_{\bf 8}g\rangle$ and $|(^3S_1)_{\bf 8}g\rangle$ by using the same color-octet matrix elements as those of Ref.\cite{Chang:2005bf}. }
\label{totcro}
\end{table*}

Total cross sections for the $B_c$ meson production under various hadron-hadron collision modes via the gluon-gluon fusion mechanism are presented in Table~\ref{totcro}. The production cross-sections of $B^{(*)}_c$ in the $p$-N and N-N collision modes are much larger than that of the $pp$ collision, due to the large nuclear number in the Pb nucleus. For the case of production at the LHC, the relative importance of the total cross sections under different collision modes is $\sigma_{pp}|_{\sqrt{S}=13\;{\rm TeV}}:\sigma_{p\rm Pb}|_{\sqrt{S}=8.16\;{\rm TeV}}:\sigma_{\rm PbPb}|_{\sqrt{S}=5.02\;{\rm TeV}}=1:78:8736$, where the contributions from $B_c$ and $B_c^*$ have been summed up. As required, due to the shadowing effect, this ratio is smaller than the simple estimation of $1:N_{\rm Pb}:N_{\rm Pb}^2$.

As a useful reference, we also present the simple prediction on the color-octet states $|(^1S_0)_{\bf 8}g\rangle$ and $|(^3S_1)_{\bf 8}g\rangle$ at the RHIC and the LHC in Table \ref{totcro}, in which the same color-octet matrix elements as those of Ref.~\cite{Chang:2005bf} are adopted. Table~\ref{totcro} shows that the color-octet contribution are about $2\% \sim 4\%$ of the color-singlet $S$-wave states for both the $p$-N and N-N collisions at the RHIC and the LHC. Thus in the following discussions, we shall concentrate on the color-singlet $S$-wave production~\footnote{The color-octet production cross-sections depend heavily on the magnitude of the color-octet matrix elements. The color-octet components are sizable in comparison to the color-singlet $P$-wave states, thus they should be taken into consideration when discussing the production of the $P$-wave states. A detailed discussion on the production of the $P$-wave states together with those two color-octet $S$-wave states is in preparation.}.

To estimate how many $B_c$-meson events can be produced at the RHIC and LHC, we sum up the $B_c(^1S_0)$ and $B_c^*(^3S_1)$ contributions together and adopt the designed luminosity of RHIC and LHC~\cite{Patrignani:2016xqp} to do the estimation. For convenience, we use a short notation to express the integrated luminosity per operation year~\footnote{At the RHIC, one operation year equals to $10^7$s for $p$-N and N-N collisions. At the LHC, one operation year equals to $10^7$s for $pp$-collision and $10^6$s for $p$-N and N-N collisions~\cite{Carminati:2004fp, Alessandro:2006yt}.}, i..e ${\cal L}_{\rm M}^{\rm R(L)}$, where $\rm R(L)$ stands for RHIC (LHC), and the subscript M represents the mentioned collision modes $pp$, $p$-N, and N-N, respectively.

At the RHIC, the designed luminosities for the $p$-Au and the Au-Au collision modes are $4.5\times 10^{29}\,\rm cm^{-2}s^{-1}$ and $8.0\times 10^{27}\,\rm cm^{-2}s^{-1}$, and the integrated luminosities are ${\cal L}_{p\rm Au}^{\rm R}=4.5 \,{\rm pb}^{-1}$ and ${\cal L}_{\rm AuAu}^{\rm R}=80 \,{\rm nb}^{-1}$, respectively. Consequently, we shall have $1.2 \times 10^5$ and $4.7 \times 10^5$ $B_c$-meson events to be generated in $p$-Au and Au-Au collisions at the RHIC in one operation year.

At the LHC, the designed luminosities for the $pp$, $p$-Pb and the Pb-Pb collision modes are $5.0\times 10^{33}\,\rm cm^{-2}s^{-1}$, $5.0\times 10^{29}\,\rm cm^{-2}s^{-1}$ and $3.6\times 10^{27}\,\rm cm^{-2}s^{-1}$ and the integrated luminosities are ${\cal L}_{pp}^{\rm L}=50 \,{\rm fb}^{-1}$, ${\cal L}_{p\rm Pb}^{\rm L}=0.5 \,{\rm pb}^{-1}$ and ${\cal L}_{\rm PbPb}^{\rm L}=3.6 \,{\rm nb}^{-1}$, respectively. Consequently, we shall have $7.4\times 10^9$, $5.8 \times 10^6$ and $4.6 \times 10^6$ $B_c$-meson events to be generated in $pp$, $p$-Pb and Pb-Pb collisions at the LHC in one operation year.

Thus, Table~\ref{totcro} shows that in addition to the $pp$ collision mode at the LHC, sizable $B_c$ meson events can also be produced at both the RHIC and the LHC via the $p$-N and N-N collision modes. The $p$-N and N-N collision modes should also be capable of comprehensive studies on various aspects of $B_c$-meson physics.

\subsection{Differential distributions of the $B_c (B_c^*)$-meson production via the $p$-N and N-N collision modes}

\begin{figure}[htb]
\includegraphics[width=0.48\textwidth]{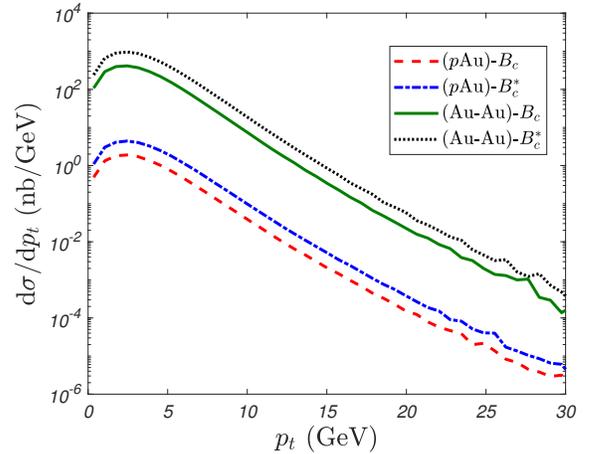}
\caption{The $p_t$-distributions of the $B_c^{(*)}$ meson production via the $p$-Au and Au-Au collision modes at the RHIC. $\sqrt{S_{p\rm Au}}=200\,\rm GeV$ and $\sqrt{S_{\rm AuAu}}=200\,\rm GeV$. }
\label{ptpr}
\end{figure}

\begin{figure}[htb]
\includegraphics[width=0.48\textwidth]{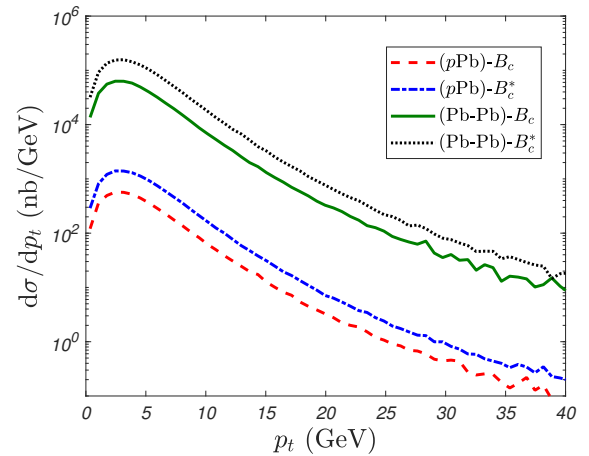}
\caption{The $p_t$-distribution of the $B_c^{(*)}$-meson production via the $p$-Pb and Pb-Pb collision modes at the LHC. $\sqrt{S_{p\rm Pb}}=8.16\,\rm TeV$ and $\sqrt{S_{\rm PbPb}}=5.02\,\rm TeV$. }
\label{ptpl}
\end{figure}

We present the $p_t$ distributions of the $B_c^{(*)}$-meson productions at the RHIC and the LHC via the $p$-N and the N-N collision modes in Figs.(\ref{ptpr}, \ref{ptpl}). The $p_t$ distributions for the $p$-N and the N-N collision modes at the RHIC and the LHC are close in shape, which shall first increase and then decreases quickly with the increment of $p_t$.

\begin{figure}[htb]
\includegraphics[width=0.48\textwidth]{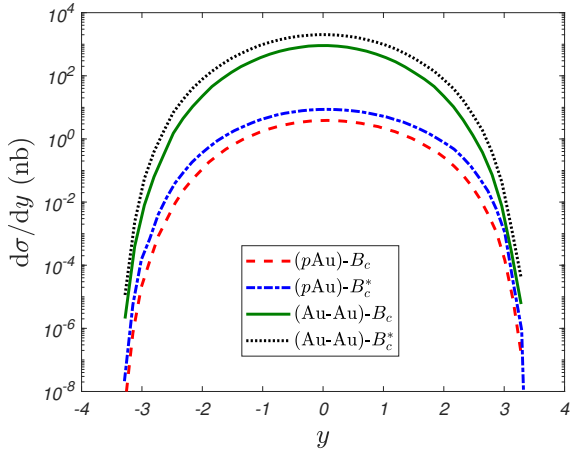}
\caption{The $y$-distribution of the $B_c^{(*)}$ meson production via the $p$-Au and Au-Au collision modes at the RHIC. $\sqrt{S_{p\rm Au}}=200\,\rm GeV$ and $\sqrt{S_{\rm AuAu}}=200\,\rm GeV$.}
\label{rapr}
\end{figure}

\begin{figure}[htb]
\includegraphics[width=0.48\textwidth]{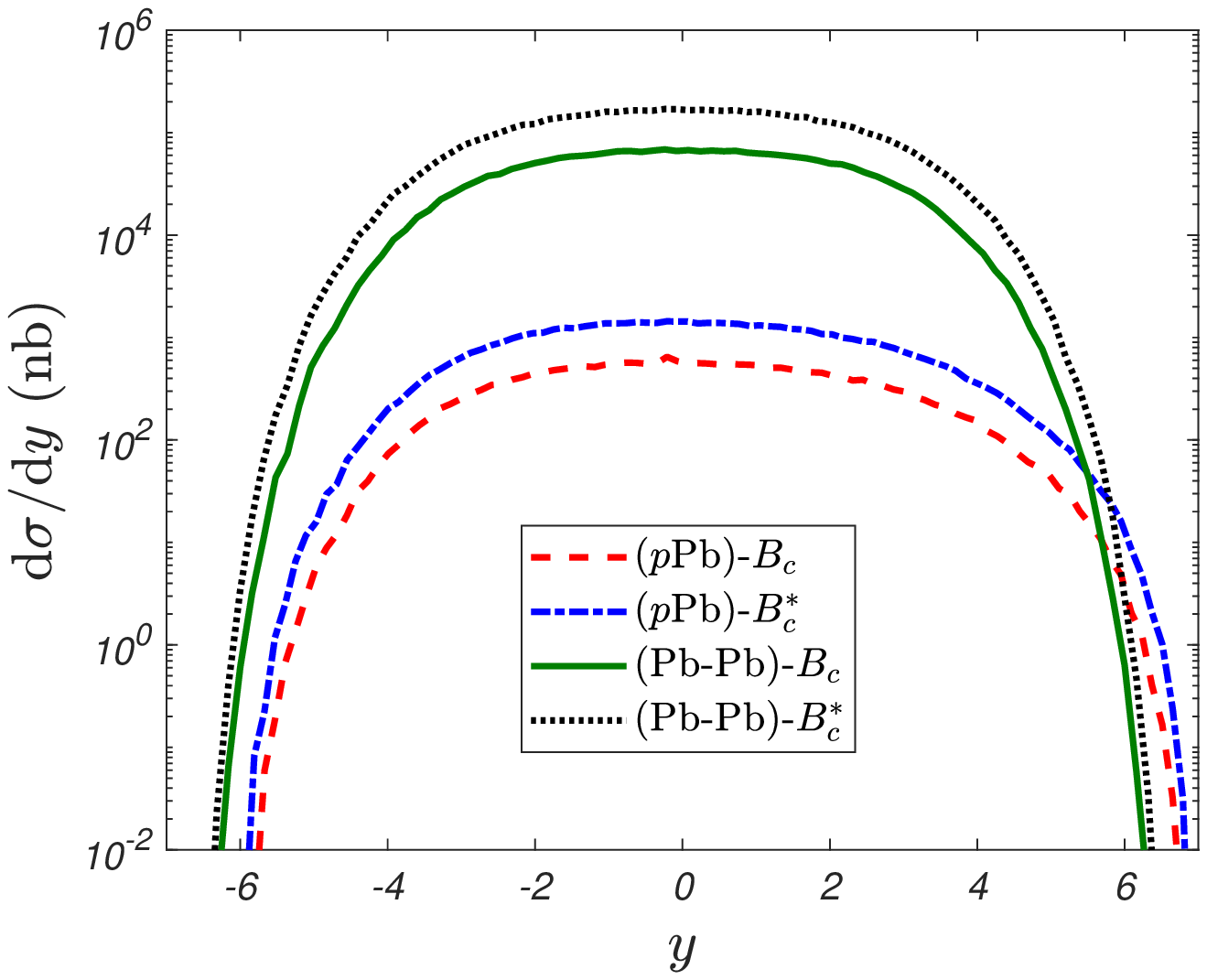}
\caption{The $y$-distribution of the $B_c^{(*)}$-meson production via the $p$-Pb and Pb-Pb collision modes at the LHC. $\sqrt{S_{p\rm Pb}}=8.16\,\rm TeV$ and $\sqrt{S_{\rm PbPb}}=5.02\,\rm TeV$.}
\label{rapl}
\end{figure}

\begin{figure}[htb]
\includegraphics[width=0.48\textwidth]{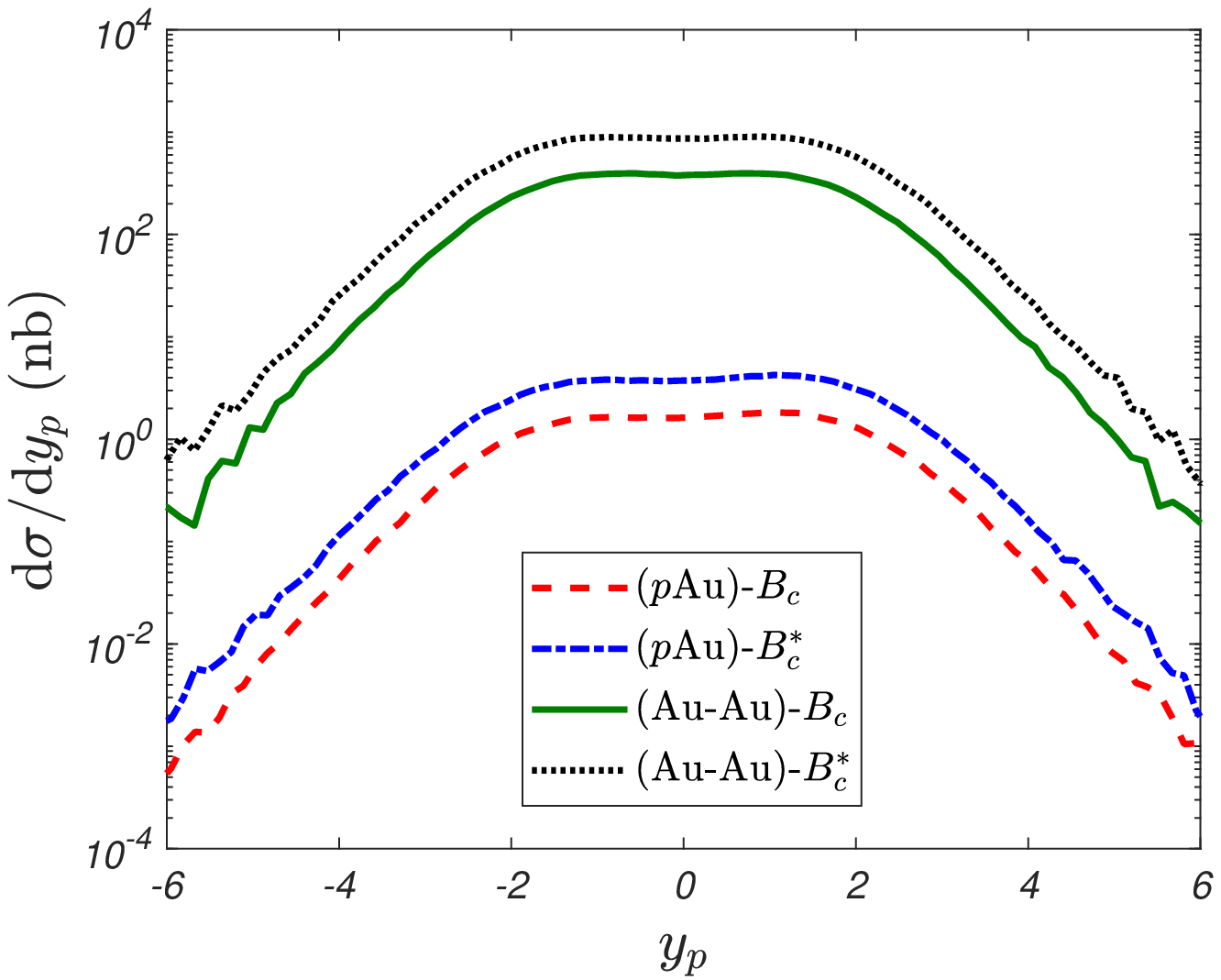}
\caption{The $y_p$-distribution of the $B_c^{(*)}$ meson production via the $p$-Au and Au-Au collision modes at the RHIC. $\sqrt{S_{p\rm Au}}=200\,\rm GeV$ and $\sqrt{S_{\rm AuAu}}=200\,\rm GeV$.}
\label{prapr}
\end{figure}

\begin{figure}[htb]
\includegraphics[width=0.48\textwidth]{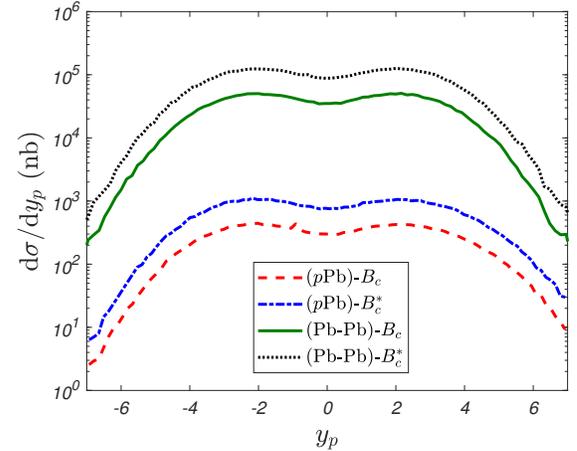}
\caption{The $y_p$-distribution of the $B_c^{(*)}$ meson production via the $p$-Pb and Pb-Pb collision modes at the LHC. $\sqrt{S_{p\rm Pb}}=8.16\,\rm TeV$ and $\sqrt{S_{\rm PbPb}}=5.02\,\rm TeV$.}
\label{prapl}
\end{figure}

We present the rapidity ($y$) and pseudo-rapidity ($y_p$) distributions of the $B_c^{(*)}$-meson productions at the RHIC and the LHC via the $p$-N and N-N collision modes in Figs.(\ref{rapr}, \ref{rapl}, \ref{prapr}, \ref{prapl}). Those distributions are asymmetric for the $p$-Au and $p$-Pb collision modes, which are more obvious for the $p$-Pb collision mode due to the fact that much more small $x$ events appearing at the LHC, thus the differences between the proton gluon PDF and the nucleus gluon PDF are greatly amplified. There are plateau for the rapidity and pseudo-rapidity distributions with $|y|\leq2$ or $|y_p|\leq2$ at the RHIC for the $p$-Au and Au-Au collision modes. At the LHC, such plateau become broader, which change to $|y|\leq4$ or $|y_p|\leq4$ for the $p$-Pb and Pb-Pb collision modes, respectively.

\begin{table*}[htb]
\begin{tabular}{|c|c|c|c|c|c|c|c|c|}
\hline
  & \multicolumn{4}{|c|}{RHIC} & \multicolumn{4}{|c|}{LHC} \\
\hline
  & \multicolumn{2}{|c|} {$p$-Au (0.2 TeV)}& \multicolumn{2}{|c|} {Au-Au (0.2 TeV)} & \multicolumn{2}{|c|} {$p$-Pb (8.16 TeV)}& \multicolumn{2}{|c|} {Pb-Pb (5.02 TeV)} \\
\hline
  & $\sigma_{B_c}$ (nb) & $\sigma_{B_c^*}$ (nb)  & $\sigma_{B_c}$ (nb) & $\sigma_{B_c^*}$ (nb) & $\sigma_{B_c}$ (nb) & $\sigma_{B_c^*}$ (nb) & $\sigma_{B_c}$ (nb) & $\sigma_{B_c^*}$ (nb)  \\
\hline
$p_t\geq2 \; \rm GeV$~& $5.82$ & $1.39\times10^1$ & $1.25\times 10^3$ & $2.98\times 10^3$ & $2.72\times 10^3$ & $6.86\times 10^3$ & $3.01\times 10^5$ & $7.62\times 10^5$  \\
\hline
$p_t\geq4 \; \rm GeV$~& $1.42$ & $5.93$ & $5.08\times10^2$ & $1.25\times 10^3$ & $1.60\times 10^3$ & $4.08\times 10^3$ & $1.75\times 10^5$ & $4.51\times 10^5$    \\
\hline
$p_t\geq6 \; \rm GeV$~& $7.58\times10^{-1}$ & $1.91$ & $1.54\times10^2$ & $3.84\times10^2$ & $8.02\times10^2$ & $2.06\times 10^3$ & $8.90\times 10^4$ & $2.26\times 10^5$  \\
\hline
\end{tabular}
\caption{Total cross sections (in unit: nb) of the $B_c^{(*)}$-meson production under various transverse momentum cuts at the RHIC and the LHC via $p$-N and N-N collision modes, respectively. }
\label{ptc}
\end{table*}

\begin{table*}[htb]
\begin{tabular}{|c|c|c|c|c|c|c|c|c|}
\hline
 & \multicolumn{4}{|c|}{RHIC} & \multicolumn{4}{|c|}{LHC} \\
\hline
  & \multicolumn{2}{|c|} {$p$-Au (0.2 TeV)}& \multicolumn{2}{|c|} {Au-Au (0.2 TeV)} & \multicolumn{2}{|c|} {$p$-Pb (8.16 TeV)}& \multicolumn{2}{|c|} {Pb-Pb (5.02 TeV)} \\
\hline
  & $\sigma_{B_c}$ (nb) & $\sigma_{B_c^*}$ (nb)  & $\sigma_{B_c}$ (nb) & $\sigma_{B_c^*}$ (nb) & $\sigma_{B_c}$ (nb) & $\sigma_{B_c^*}$ (nb) & $\sigma_{B_c}$ (nb) & $\sigma_{B_c^*}$ (nb)  \\
\hline
$|y|\leq1$~& $6.31$ & $1.44\times10^1$ & $1.44\times 10^3$ & $3.26\times 10^3$ & $1.10\times 10^3$ & $2.75\times 10^3$ & $1.30\times 10^5$ & $3.32\times 10^5$ \\
\hline
$|y|\leq2$~& $8.09$ & $1.90\times10^1$ & $1.75\times 10^3$ & $4.11\times 10^3$ & $2.09\times 10^3$ & $5.12\times 10^3$ & $2.46\times 10^5$ & $6.12\times 10^5$ \\
\hline
$|y|\leq3$~& $8.17$ & $1.93\times10^1$ & $1.76\times 10^3$ & $4.15\times 10^3$ & $2.81\times 10^3$ & $6.96\times 10^3$ & $3.27\times 10^5$ & $8.19\times 10^5$ \\
\hline
\end{tabular}
\caption{Total cross sections (in unit: nb) of the $B_c^{(*)}$-meson production under various rapidity cuts at the RHIC and the LHC via $p$-N and N-N collision modes, respectively.}
\label{yc}
\end{table*}

In a high-energy collider, the $B_c$ meson events with a small $p_t$ and/or a large rapidity $y$, indicating they are moving close to the beam direction, cannot be detected by the detectors directly, so those events cannot be utilized for experimental studies in common cases. Considering the detectors¡¯ abilities and in order to offer experimental references, we try various cuts accordingly in the estimate of the $B_c$ meson production.

We present the cross sections under typical cuts $p_t\geq2\,\rm GeV$, $p_t\geq4\,\rm GeV$, and $p_t\geq6\,\rm GeV$ in Table~\ref{ptc}. If taking the $p_{t}$ cut equals to $2$ GeV ($6$ GeV), the total cross section shall be reduced by $\sim17\%$ ($\sim75\%$) for both the $p$-Pb  and Pb-Pb collision modes at the LHC, and by $\sim28\%$ ($\sim90\%$) for both the $p$-Au and Au-Au collision modes at the RHIC. Similarly, we present the cross sections for three typical rapidity cuts,  $|y|\leq1$, $|y|\leq2$, and $|y|\leq3$, in Table~\ref{yc}.

\subsection{Uncertainties from different choices of the heavy quark masses and renormalization scale}

\begin{table*}[htb]
\begin{tabular}{|c|c|c|c|c|c|c|c|c|}
\hline
  & \multicolumn{4}{|c|}{RHIC} & \multicolumn{4}{|c|}{LHC} \\
\hline
  & \multicolumn{2}{|c|} {$p$-Au (0.2 TeV)}& \multicolumn{2}{|c|} {Au-Au (0.2 TeV)} & \multicolumn{2}{|c|} {$p$-Pb (8.16 TeV)}& \multicolumn{2}{|c|} {Pb-Pb (5.02 TeV)} \\
\hline
$m_c$ & $\sigma_{B_c}$ (nb) & $\sigma_{B_c^*}$ (nb)  & $\sigma_{B_c}$ (nb) & $\sigma_{B_c^*}$ (nb) & $\sigma_{B_c}$ (nb) & $\sigma_{B_c^*}$ (nb) & $\sigma_{B_c}$ (nb) & $\sigma_{B_c^*}$ (nb)  \\
\hline
1.4 GeV~& $1.06\times10^1$ & $2.57\times10^1$ & $2.29\times 10^3$ & $5.55\times 10^3$ & $4.05\times 10^3$ & $1.05\times 10^4$ & $4.53\times 10^5$ & $1.15\times 10^6$ \\
\hline
1.5 GeV~& $8.19$ & $1.93\times10^1$ & $1.76\times 10^3$ & $4.15\times 10^3$ & $3.29\times 10^3$ & $8.26\times 10^3$ & $3.69\times 10^5$ & $9.21\times 10^5$ \\
\hline
1.6 GeV~& $6.42$ & $1.47\times10^1$ & $1.38\times 10^3$ & $3.14\times 10^3$ & $2.75\times 10^3$ & $6.69\times 10^3$ & $3.04\times 10^5$ & $7.44\times 10^5$ \\
\hline
\end{tabular}
\caption{Total cross sections (in unit: nb) of the $B_c^{(*)}$-meson production with different $c$-quark masses at the RHIC and the LHC via $p$-N and N-N collision modes, respectively. }
\label{mcu}
\end{table*}

\begin{table*}[htb]
\begin{tabular}{|c|c|c|c|c|c|c|c|c|}
\hline
  & \multicolumn{4}{|c|}{RHIC} & \multicolumn{4}{|c|}{LHC} \\
\hline
  & \multicolumn{2}{|c|} {$p$-Au (0.2 TeV)} & \multicolumn{2}{|c|} {Au-Au (0.2 TeV)} & \multicolumn{2}{|c|} {$p$-Pb (8.16 TeV)}& \multicolumn{2}{|c|} {Pb-Pb (5.02 TeV)} \\
\hline
$m_b$ & $\sigma_{B_c}$ (nb) & $\sigma_{B_c^*}$ (nb)  & $\sigma_{B_c}$ (nb) & $\sigma_{B_c^*}$ (nb) & $\sigma_{B_c}$ (nb) & $\sigma_{B_c^*}$ (nb) & $\sigma_{B_c}$ (nb) & $\sigma_{B_c^*}$ (nb)  \\
\hline
4.7 GeV~& $1.03\times10^1$ & $2.38\times10^1$ & $2.23\times 10^3$ &  $5.14\times 10^3$ & $3.83\times 10^3$ & $9.40\times 10^3$ & $4.26\times 10^5$ & $1.04\times 10^6$ \\
\hline
4.9 GeV~& $8.19$ & $1.93\times10^1$ & $1.76\times 10^3$ & $4.15\times 10^3$ & $3.29\times 10^3$ & $8.26\times 10^3$ & $3.69\times 10^5$ & $9.21\times 10^5$ \\
\hline
5.1 GeV~& $6.55$ & $1.57\times10^1$ & $1.40\times 10^3$ & $3.35\times 10^3$ & $2.86\times 10^3$ & $7.29\times 10^3$ & $3.19\times 10^5$ & $8.12\times 10^5$ \\
\hline
\end{tabular}
\caption{Total cross sections (in unit: nb) of the $B_c^{(*)}$-meson production with different $b$-quark masses at the RHIC and the LHC via $p$-N and N-N collision modes, respectively. }
\label{mbu}
\end{table*}

In this subsection, we discuss the uncertainties from different choice of heavy quark mass and the renormalization scale.

To estimate the uncertainties from the heavy quark masses $m_c$ and $m_b$, we take $m_c=1.50\pm0.10$ GeV and $m_b=4.9\pm0.20$ GeV. When the quark masses are changed, the mass of the $B_c$ meson shall be alternated accordingly to ensure $M_{B_c}=m_b+m_c$. We present the results in Tables \ref{mcu} and \ref{mbu}. Those two tables show that the total cross section depends heavily on the value of the $c$-quark or the $b$-quark mass, which decreases with the increment of the quark mass and is more sensitive to the $c$-quark mass. By summing the spin-singlet and spin-triplet contributions together, Table~\ref{mcu} shows that by taking $\Delta m_c=\pm0.1$ GeV, the cross section shall be changed by $[-20\%, +34\%]$ for the RHIC and by $[-17\%, 27\%]$ for the LHC; Table~\ref{mbu} shows that by taking $\Delta m_b=\pm0.2$ GeV, the total cross section shall be changed by $[-15\%, +26\%]$ for the RHIC and by $[-11\%, +16\%]$ for the LHC.

\begin{table*}[htb]
\begin{tabular}{|c|c|c|c|c|c|c|c|c|}
\hline
 & \multicolumn{4}{|c|}{RHIC} & \multicolumn{4}{|c|}{LHC} \\
\hline
 & \multicolumn{2}{|c|} {$p$-Au (0.2 TeV)}& \multicolumn{2}{|c|} {Au-Au (0.2 TeV)} & \multicolumn{2}{|c|} {$p$-Pb (8.16 TeV)}& \multicolumn{2}{|c|} {Pb-Pb (5.02 TeV)} \\
\hline
$\mu_R$ & $\sigma_{B_c}$ (nb) & $\sigma_{B_c^*}$ (nb)  & $\sigma_{B_c}$ (nb) & $\sigma_{B_c^*}$ (nb) & $\sigma_{B_c}$ (nb) & $\sigma_{B_c^*}$ (nb) & $\sigma_{B_c}$ (nb) & $\sigma_{B_c^*}$ (nb)  \\
\hline
$\sqrt{\hat{s}}$~& 2.32 & 5.34 & $4.95\times10^2$ & $1.12\times10^3$ & $1.83\times10^3$ & $4.51\times10^3$ & $1.94\times10^5$ & $4.77\times10^5$ \\
\hline
$\sqrt{\hat{s}}/2$~& 5.29 & $1.21\times 10^1$ & $1.14\times 10^3$ & $2.61\times10^3$ & $2.46\times10^3$ & $6.06\times10^3$ & $2.69\times10^5$ & $6.58\times10^5$ \\
\hline
$M_t$~& 8.19 & $1.93\times10^1$ & $1.76\times 10^3$ & $4.15\times10^3$ & $3.29\times10^3$ & $8.26\times10^3$ & $3.69\times10^5$ & $9.21\times10^5$ \\
\hline
\end{tabular}
\caption{Total cross sections (in unit: nb) of $B_c^{(*)}$-meson production with three typical choice of renormalization scale $\mu_R$ at RHIC and LHC via $p$-N and N-N collision modes, respectively.}
\label{scu}
\end{table*}

As an estimation of renormalization scale dependence, we take three typical scales to calculate the total cross section. The results are presented in Table~\ref{scu}. In addition to the transverse mass $M_t$, we take another two choices $\mu_R=\sqrt{\hat{s}}/2$ and $\mu_R=\sqrt{\hat{s}}$ to do the estimation, where $\sqrt{\hat{s}}$ is the center-of-mass energy of subprocess.

At the tree level, there is large scale uncertainty. By summing the spin-singlet and spin-triplet contributions together, the uncertainties are $\sim70\%$ for the $p$-Au and Au-Au collision modes at the RHIC, which change down to $\sim 45\%$ for the $p$-Pb and Pb-Pb collision modes at the LHC. To suppress the scale uncertainty, it is helpful to finish a next-to-leading order calculation. In fact, even if we have finished such a high-order calculation, we still need a proper scale-setting approach such that to achieve a scheme-and-scale independent prediction at lower orders~\cite{Wu:2013ei, Wu:2014iba}.

\section{Summary} \label{summary}

By taking the gluon-gluon fusion production mechanism into consideration, we have performed a detailed discussion on the $B_c (B_c^*)$-meson production via the $p$-N and N-N collision modes at the RHIC and LHC, respectively. It is found that sizable number of $B_c$-meson events may be produced at the RHIC and LHC via the $p$-N and N-N collision modes. For instance, if assuming all the spin-triplet $B^*_c$ meson decay to the spin-singlet $B_c$ meson with $100\%$ probability, $1.2 \times 10^5$ and $4.7 \times 10^5$ $B_c$-meson events shall be produced via the $p$-Au and Au-Au collision modes at the RHIC in one operation year; and $5.8 \times 10^6$ and $4.6 \times 10^6$ $B_c$-meson events shall be produced via the $p$-Pb and Pb-Pb collision modes at the LHC in one operation year.

Differential distributions for various collision modes have been presented in Figs.(\ref{ptpr}, \ref{ptpl}, \ref{rapr}, \ref{rapl}, \ref{prapr}, \ref{prapl}). The $p_t$ distributions are close in shape to each other, which decrease quickly in large $p_t$ region. The rapidity and pseudo-rapidity distributions in the $p$-N collision mode are asymmetric in certain manner, which become more and more obvious at the LHC due to more small $x$-events appear and the differences between the gluon components in proton and in nucleus shall be greatly amplified. There are plateau in the rapidity and pseudo-rapidity distributions for the production of the $B_c (B_c^*)$-meson via various heavy-ion collision modes. Those plateau become broader and broader with increasing collision energies among the incident hadrons, e.g. at the RHIC the plateau appear within the region of $|y|\leq2$ or $|y_p|\leq2$, which change to $|y|\leq4$ or $|y_p|\leq4$ for the case of LHC.

There are large uncertainties from the value of the heavy quark masses and the chosen renormalization scale, which could be suppressed by including high-order terms.

In summary for the productions of $B_c (B_c^*)$-meson at the RHIC, we obtain
\begin{eqnarray}\label{fin1}
\sigma_{B_c}^{p\rm Au} &=& 8.19^{+2.39+2.11}_{-1.77-1.64}  \;{\rm nb}, \\
\sigma_{B_c^*}^{p\rm Au} &=& 19.3^{+6.35+4.51}_{-4.65-3.62}  \;{\rm nb},  \\
\sigma_{B_{c}}^{\rm AuAu} &=& 1.76^{+0.52+0.47}_{-0.38-0.36}  \;{\rm  \mu b}, \\
\sigma_{B_{c}^*}^{\rm AuAu} &=& 4.15^{+1.41+0.99}_{-1.01-0.79} \;{\rm \mu b};
\end{eqnarray}
and for the production at the LHC, we obtain
\begin{eqnarray}\label{fin2}
\sigma_{B_c}^{p\rm Pb} &=& 3.29^{+0.76+0.54}_{-0.55-0.44}\;{\rm \mu b}, \\
\sigma_{B_c^*}^{p\rm Pb} &=& 8.26^{+2.22+1.14}_{-1.57-0.97}\;{\rm \mu b}, \\
\sigma_{B_{c}}^{\rm PbPb} &=& (3.69\times10^2)^{+8.35\times10^1+5.69\times10^1}_{-6.49\times10^1-5.01\times10^1}\;{\rm \mu b}, \\
\sigma_{B_{c}^*}^{\rm PbPb} &=& (9.21\times10^2)^{+2.31\times10^2+1.22\times10^2}_{-1.78\times10^2-1.09\times10^2}\;{\rm \mu b}.
\end{eqnarray}
Here the first error is for $\Delta m_c=\pm0.1$ GeV and the second error is for $\Delta m_b=\pm0.2$ GeV, and no $p_t$ and rapidity cuts are taken in those predictions.

We have shown that sizable number of $B_c$ meson events can be produced at both the RHIC and LHC via the $p$-N and N-N collision modes. On one hand, it indicates that one may study the $B_c$-meson properties by using more collision modes other than the usually considered $pp$ collision mode. On the other hand, it shows that to study the QGP via the signals of the $B_c$ meson is feasible. Moreover the information shall be useful for testing the nucleus PDFs, especially to test the shadowing effects etc within the nucleus. At present our discussions are concentrated on the $S$-wave $(c\bar{b})$-quarkonium states and on the dominant direct gluon-gluon fusion production mechanism, the studies on the production of higher excited $(c\bar{b})$-quarkonium states and on different production mechanisms, such as intrinsic charm or extrinsic charm mechanism, are also important. \\

{\bf Acknowledgement:} We would like to thank Hua-Yong Han for helpful discussions. This work was supported in part by Natural Science Foundation of China under Grant No.11605029, No.11675239, No.11625520, and No.11535002, and Science project of Colleges and universities directly under the Guangzhou Education Bureau under Grant No. 1201630158, and The Foundation for Fostering the Scientific and Technical Innovation of Guangzhou University.


\end{document}